\documentclass[aps,prd,twocolumn,superscriptaddress,showpacs]{revtex4}
\usepackage{amsmath}
\usepackage{amssymb}
\usepackage{graphicx}
\newcommand{\bastar}{\begin{eqnarray*}}
\newcommand{\eastar}{\end{eqnarray*}}
\relax

\newcommand{\bea}{\begin{eqnarray}}
\newcommand{\eea}{\end{eqnarray}}
\newcommand{\be}{\begin{equation}}
\newcommand{\ee}{\end{equation}}

\begin{document}
\title{Problem of time and Hamiltonian reduction in the (2+2)\ formalism}
\author{J.H. Yoon}
\email{yoonjh@konkuk.ac.kr}
\affiliation{School of Physics,
Konkuk University, Seoul 143-701, Korea}

\begin{abstract}
We apply the Hamiltonian reduction procedure to general spacetimes of 4-dimensions
in the (2+2) formalism and find privileged spacetime coordinates in which the
physical Hamiltonian is expressed in true degrees of freedom only, namely,
the conformal two-metric on the
cross section of null hypersurfaces and its conjugate momentum.
The physical time is the area element of the cross section of
null hypersurface, and the physical radial coordinate is defined by 
{\it equipotential} surfaces on a given spacelike hypersurface of constant physical time.
The physical Hamiltonian is {\it constraint-free} and manifestly {\it positive-definite}
in the privileged coordinates.
We present the complete set of the Hamilton's equations, and
find that they coincide with the Einstein's equations written in the privileged 
coordinates. This shows that our Hamiltonian reduction 
is self-consistent and respects the general covariance. 
This work is a generalization of ADM Hamiltonian reduction of midi-superspace to 
4-dimensional spacetimes with no isometries.

\end{abstract}
\pacs{04.20.Fy, 04.50.Cd, 04.60.Ds, 02.20Tw}

\maketitle

It has been known for a long time
that true degrees of freedom of general relativity reside in the
conformal two-metric of the cross section of null 
hypersurfaces\cite{sachs62,dinverno78}. 
Eliminating unphysical degrees of freedom by identifying 
arbitrary
spacetime coordinates with certain functions in phase space
and thereby presenting the theory in terms of physical degrees of freedom only
in the privileged coordinates, free from constraints,
is known as ADM Hamiltonian reduction\cite{ADM,rovelli94,brown95}. 
K. Kucha{\v r} applied this procedure to spacetimes that
admit two commuting Killing vector fields, known as
midi-superspace\cite{kuchar71,husain94,torre96}, and showed that
the Einstein's theory is equivalent to cylindrical massless scalar field theory
propagating
in the 1+1 dimensional Minkowski spacetime.

In this Letter, we apply ADM-like Hamiltonian reduction procedure
to general spacetimes of 4-dimensions using the (2+2)
decomposition\cite{yoon92,yoon93a,yoon99a,yoon99c,yoon04}. 
The area element of the cross section of null hypersurfaces emerges as the physical time, and the physical radial coordinate is defined by {\it equipotential} 
surfaces on a given spacelike hypersurface of constant physical time.
We present the fully reduced physical Hamiltonian in these privileged coordinates\cite{husain12},
which turns out to be positive-definite. The momentum constraints 
are simply the {\it defining} equations 
of the momenta of the theory in term of physical degrees of freedoms\cite{kuchar71}; 
hence they are no longer constraints and
the theory becomes constraint-free. Moreover,  we find that our Hamiltonian reduction is self-consistent because the Hamilton's equations
of motion obtained through this Hamiltonian reduction are {\it identical} to the
Ricci-flat equations in the privileged coordinates.
As a by-product of this Hamiltonian reduction, we found 
an independent proof of {\it topological censorship}\cite{hawking72,fsw93,chru94,jacobson95,galloway96,gsww99}, 
which follows directly from one of the 
Einstein's equations in these coordinates.

Let us recall that the metric in the (2+2) decomposition\cite{sachs62,dinverno78,yoon92,yoon93a,yoon99a,yoon99c,yoon04} of 
4-dimensional spacetimes can be written as
\bea 
& & \hspace{-0.6cm}
ds^2 = 2dudv - 2hdu^2 +\tau\rho_{ab}
 \left( dy^a + A_{+}^{\ a} du + A_{-}^{\ a} dv \right)  \nonumber\\
& & \hspace{0.2cm} 
\times
\left( dy^b + A_{+}^{\ b} du + A_{-}^{\ b} dv  \right).          \label{yoon}
\eea
The vector fields
$\hat{\partial}_{+}$ and $\hat{\partial}_{-}$ defined as
\be
\hat{\partial}_{+}:={\partial \over \partial u} - A_{+}^{\ a} {\partial \over \partial y^{a}},
\hspace{0.2cm}
\hat{\partial}_{-}:={\partial \over \partial v} - A_{-}^{\ a} {\partial \over \partial y^{a}}
\ \ (a=2,3)
\ee
are horizontal vector fields orthogonal to the two-dimensional spacelike surface
$N_{2}$ generated by $\partial_{a} = {\partial / \partial y^{a}}$. 
The metric on $N_{2}$  is given by
$\tau\rho_{ab}$ (with ${\rm det}\rho_{ab}=1$), and $\hat{\partial}_{-}$
is a null vector field with a zero norm.
In this Letter, we choose the sign $-2h>0$ so that $v= {\rm constant}$ hypersurface
is spacelike.

As was shown in \cite{yoon04}, the Einstein's equations
can be obtained from the variational principle of the following action integral,
\bea
& & \hspace{-0.4cm}
S=\int \!\! dv du d^{2} y \{ \pi_{\tau}\dot{\tau} + \pi_{h}\dot{h}
+ \pi_{a}\dot{A}_{+}^{\ a}    
+ \pi^{ab}\dot{\rho}_{ab}   \nonumber\\
& & 
 \hspace{0.5cm}
  -``1" \cdot C -``0" \cdot C_{+} - A_{-}^{\ a}C_{a} \},                              \label{bareaction}
\eea
where the overdot $\dot{} = \partial_{-}$, and
$``1"$, $``0"$, and $A_{-}^{\ a}$ are Lagrange multipliers that enforce
the constraints $C$, $C_{+}$,  and $C_{a}$, which are given by
\begin{eqnarray}
& & \hspace{-0.6cm}  ({\rm i}) \
C:=  {1\over 2}\pi_{h}\pi_{\tau}
-{h\over 4\tau}\pi_{h}^{2}
-{1\over 2\tau}\pi_{h}D_{+}\tau
+{1\over 2\tau^{2}}\rho^{a b}\pi_{a}\pi_{b}\nonumber\\
& &
-{\tau\over 8h}\rho^{a b} \rho^{c d}
(D_{+}\rho_{a c}) (D_{+}\rho_{b d})
-{1\over 2h \tau}
\rho_{a b}\rho_{c d}\pi^{a c}\pi^{b d}  \nonumber\\
& &
-{1\over 2h}\pi^{a c}D_{+}\rho_{a c}
-\tau R_{(2)}  + D_{+}\pi_{h}
- \partial_{a}(\tau^{-1}\rho^{a b} \pi_{b}) \nonumber\\
& & =0,  \label{C}
\end{eqnarray}
\begin{eqnarray}
& & \hspace{-0.6cm} ({\rm ii}) \ 
C_{+}:=\pi_{\tau}D_{+}\tau + \pi_{h}D_{+}h + \pi^{a b}D_{+}\rho_{a b} \nonumber\\
& & - 2D_{+}(
h \pi_{h}  +  D_{+} \tau )
+2\partial_{a}(h \tau^{-1}\rho^{a b}\pi_{b}
 +\rho^{a b}\partial_{b}h )\nonumber\\
& & =0,         \label{C+}
\end{eqnarray}
\begin{eqnarray}
& & \hspace{-0.6cm}({\rm iii}) \
C_{a}:=\pi_{\tau}\partial_{a}\tau +\pi_{h}\partial_{a}h
+\pi^{b c}\partial_{a}\rho_{b c}
-2\partial_{b}( \rho_{ac}\pi^{bc})  \nonumber\\
& & -D_{+}\pi_{a} - \partial_{a}(\tau \pi_{\tau}) =0. \label{CA}
\end{eqnarray}
Notice minor changes of sign from \cite{yoon04}.
Here $R_{(2)}$ is the scalar curvature of $N_{2}$,
and the diff$N_{2}$-covariant derivative\cite{yoon04} of a tensor density
$q_{a b }$ with weight $s$ is defined as
\bea
& & \hspace{-0.4cm}
D_{+}q_{a b}:= \partial_{u}q_{a b}
-[A_{+}, \ q]_{{\rm L}a b}  
= \partial_{u }q_{a b} - A_{+}^{\ c}\partial_{c}q_{ab} \nonumber\\
& &
-q_{cb}\partial_{a}A_{+}^{\ c}
-q_{ac}\partial_{b}A_{+}^{\ c}    
-s (\partial_{c}A_{+}^{\ c})q_{ab},
            \label{covdiff}
\eea
where $[A_{+}, \ q]_{{\rm L} a b}$ is the Lie
derivative of $q_{ab}$ along $A_{+}:=A_{+}^{\ a}\partial_{a}$.
Let us define a {\it potential} function $R$ and rename $A_{+}^{\ a}$ as $w^{a}$ such that
\be
D_{+}R:=-h\pi_{h},   \hspace{0.2cm}
w^{a}=A_{+}^{\ a}.  \label{iden}
\ee
If we impose the constraint $C_{+}=0$ and $C_{a}=0$,  then
the action (\ref{bareaction}) becomes
\bea
& & S=\int \!\! dv du d^{2} y \{ \pi_{\tau}\dot\tau + \pi_{R}\dot R
+ \pi_{a}^{w}\dot w^{a} + \pi^{ab}\dot\rho_{ab} -C \}           \nonumber\\
& & \ + \ {\rm total \ derivatives},   \label{action1}
\eea
where the momenta $\pi_{R}$ and ${\pi}_{a}^{w}$ conjugate to $R$ and $w^{a}$
are given by
\bea
& & \pi_{R}=-D_{+}{\ln}(-h),        \label{pir}\\
& & {\pi}_{a}^{w}=\pi_{a} - (\partial_{a} R) \ {\ln}(-h),  \label{pota}
\eea
respectively. The transformation from  $\{h, \pi_{h}; A_{+}^{\ a}, \pi_{a}\}$ to 
$\{ R, \pi_{R}; w^{a}, \pi_{a}^{w}  \}$ is clearly a {\it canonical} transformation, as it changes
the action integral by total derivatives only.
The Hamiltonian constraint in new variables becomes
\bea
& &
C=-{ 1\over 2h}\pi_{\tau} D_{+}R
- {1\over 4h \tau}(D_{+}R)^{2}  +{1\over 2h\tau}(D_{+}\tau)(D_{+}R) \nonumber\\
& & -{1\over h}D_{+}^{2}R -{1\over h}\pi_{R}D_{+}R 
 -\tau R_{(2)} \nonumber\\
& &
+{1\over 2\tau^{2}}\rho^{ab}\{ \pi_{a}^{w} +  {\ln}(-h)\partial_{a}R\}
\{ \pi_{b}^{w}  + {\ln}(-h)\partial_{b}R\} \nonumber\\
& & -{1\over 2h \tau}
\rho_{a b}\rho_{c d}\pi^{a c}\pi^{b d}
-{\tau\over 8h}\rho^{a b} \rho^{c d}
(D_{+}\rho_{a c}) (D_{+}\rho_{b d}) \nonumber\\
& &
-{1\over 2h}\pi^{a c}D_{+}\rho_{a c}
-\partial_{a}[ \tau^{-1}\rho^{ab}\{ \pi_{b}^{w}  + {\ln}(-h)\partial_{b}R \} ]        \nonumber\\
& & =0.                   \label{newham}
\eea 
{\it  Hamiltonian reduction I: Choose $\tau$ as  physical time} \\
Notice that the first term in (\ref{newham}) is linear in $\pi_{\tau}$, and all the remaining terms are
independent of  $\pi_{\tau}$,  since
$h$ depends on $\{\pi_{R}, w^{a}\}$ only, as is obvious 
from the equation (\ref{pir}). Thus, the equation of motion of $\tau$ is given by
\be
\dot{\tau}=\int \!\! du d^{2}y  {\delta C \over \delta \pi_{\tau}}
=-{1\over 2h}D_{+}R. \label{taueom}
\ee
Now, recall that $\tau=\tau(v,u,y^{a})$. Solving this equation for $v$, one may view $v$ as a function of $(\tau,u,y^{a})$  and consequently 
regard $\{ R$, $w^{a}$, $\rho_{ab} \}$ as functions of  
$(\tau,u,y^{a})$. Therefore, it follows that
\bea
& & \dot{R}=\dot{\tau}\partial_{\tau}R,  \ \dot{w}^{a}=\dot{\tau}\partial_{\tau}w^{a},  \
\dot{\rho}_{ab}=\dot{\tau}\partial_{\tau}\rho_{ab},    \nonumber\\
& & C=-({2h \over D_{+}R})\dot{\tau} C,
\eea
since ${\partial u /\partial v}={\partial y^{a} /\partial v}=0$.
Then the action (\ref{action1}) becomes
\bea
& &  \hspace{-1cm}
S=\int \!\! dvdu d^{2} y \dot{\tau} \{ 
\pi_{\tau}  +\pi_{R}\partial_{\tau} R +  \pi_{a}^{w}\partial_{\tau} w^{a}  \nonumber\\
& & 
+ \pi^{ab}\partial_{\tau}\rho_{ab} +({2h \over D_{+}R})C \}  \nonumber\\
& &  \hspace{-1cm}
=\int \!\! d\tau du d^{2} y \{
\pi_{R}\partial_{\tau} R +  \pi_{a}^{w}\partial_{\tau} w^{a}
+ \pi^{ab}\partial_{\tau}\rho_{ab}
-C^{(1)} \},  \label{action2}
\eea
where we replaced $dv \dot{\tau}$ by $d\tau$ in the second line, 
and $C^{(1)}$ is defined as
\be
C^{(1)}=- ({2h\over D_{+}R})C - \pi_{\tau}.
\ee
{\it Hamiltonian reduction II:  Choose $R$ as physical radius and fix arbitrary 
coordinates $y^{a}$ on $R={\it constant}$ subspace such that $w^{a}=0$}\\
The second step in the Hamiltonian reduction consists of {\it identifying}
arbitrary coordinate $u$ as $R$ and {\it choosing} $y^{a}=Y^{a}$ such that  the ``shift" vector $w^{a}$ on $R={\rm constant}$ subspace is zero,
\be
w^{a}=0.                         \label{zerow}
\ee
Then, it follows from (\ref{taueom}) that
\be
\dot{\tau}=-{1\over 2h} \geq 0,
\ee
which means that $\tau$ increases monotonically along the out-going 
null vector field. Thus, $\{R, Y^{a}\}$  are the privileged coordinates on  
$\tau= {\rm constant}$ hypersurface $\Sigma_{3}$,  and therefore the following equations are
trivially true,
\be
\partial_{a} \tau=\partial_{R} \tau=0.    \label{partau}
\ee
%Figure1
\begin{figure}
\vspace{-1.8in}
\hspace{-2in}
\begin{center}
\includegraphics[width=5in]{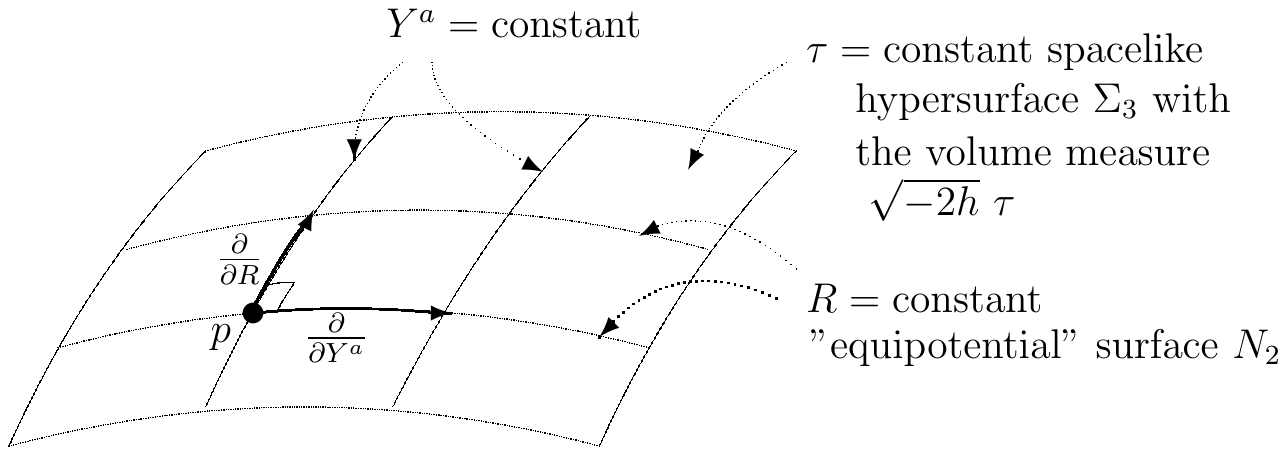}
\vspace{-4.4in}
\caption{\label{figure1} On $R={\rm constant}$ ``equipotential" surface $N_{2}$
on $\Sigma_{3}$,  $Y^{a}$ are introduced such that 
$Y^{a}={\rm constant}$ is
normal to $N_{2}$ at each point $p \in N_{2}$. Then the ``shift" vector $w^{a}$ becomes 
zero at $p$.}
\end{center}
\vspace{-0.5cm}
\end{figure}
But recall that $\tau$ is a scalar density with weight $1$, rather than a scalar function,
with respect to diff$N_{2}$ transformations. Thus,  
the covariant derivative $D_{R}\tau $ is given by
\be
D_{R}\tau = \partial_{R} \tau - w^{a}\partial_{a} \tau -( \partial_{a}w^{a} )\tau.
\ee
By the condition (\ref{zerow}) and (\ref{partau}),  $\tau$ is covariantly constant
on $\Sigma_{3}$,
\be
D_{R}\tau =0,
\ee 
which means that our choice of $\tau$ as a physical time 
is consistently defined over $\Sigma_{3}$, even though  $\tau$
is a scalar density rather than a scalar function. 
The Hamilton's equations of motion follows from the variational principle
of the action integral (\ref{action2}) 
\bea
& & \partial_{\tau}q^{I}=\int_{\Sigma_{3}} \!\!\!\! d u d^{2}y
{\delta C^{(1)} \over \delta \pi_{I}}|_{u=R, y^{a}=Y^{a} } ,  \\
& & 
\partial_{\tau}\pi_{I}=-\int_{\Sigma_{3}} \!\!\!\! d u d^{2}y
{\delta C^{(1)} \over \delta q^{I}}|_{u=R, y^{a}=Y^{a}  } ,
\eea
for $q^{I}=\{  R, w^{a},  \rho_{ab}\} $, and 
$\pi_{I}=\{ \pi_{R}, \pi_{a}^{w},  \pi^{ab} \}$.
In the following we present the main results of this Letter without derivation: \\
{\it Einstein's evolution equations}
\bea
& & 
1. {\partial R \over \partial \tau}=0 \Rightarrow  \nonumber\\
& & 
\tau R_{(2)}={1\over 2}\tau^{-2}\rho^{ab}\pi_{a}^{w}
\pi_{b}^{w}-\partial_{a}(\tau^{-1}\rho^{ab}\pi_{b}^{w})   \label{topo} \\
& & \hspace{0.1cm} ( {\rm topological \  censorship})        \nonumber\\
& & 
2. {\partial w^{a} \over \partial \tau}=0 \Rightarrow  \nonumber\\
& & \tau^{-1}\pi_{a}^{w}=-\partial_{a}{\ln}(-h)  \label{parwa}  
\hspace{0.2cm}  ({\rm superpotential})  \\ 
& & 
3. {\partial \tau\over \partial \tau}=1\  ({ \rm trivial}) \\
& & 
4.  {\partial \ {\ln}(-h) \over \partial \tau}=H_{*} - {1\over \tau}
\ ({\rm superpotential}) \\
& & 
5.  {\partial \pi_{a}^{w}\over \partial \tau}=2\tau^{-1}\pi_{a}^{w} +
(\pi^{bc}+ {1\over 2}\rho^{bd}\rho^{ce}\partial_{R}\rho_{de})\partial_{a}\rho_{bc}\nonumber\\
& & \hspace{0.2cm}
-\partial_{b}(2\pi^{bc}\rho_{ac}+\rho^{bc}\partial_{R}\rho_{ac}) \label{parpi}\\
& &
\ ( {\rm evolution \ equation \ of} \ \pi_{a}^{w})  \nonumber\\
& &  
6.   {\partial \pi_{\tau}\over \partial \tau}= {1\over 2}\tau^{-2} 
+ \tau^{-2}\rho_{ab}\rho_{cd}\pi^{ac}\pi^{bd} \nonumber\\
& & 
-{1\over 4}\rho^{ab}\rho^{cd}(\partial_{R}\rho_{ac})(\partial_{R}\rho_{bd}) 
-2\tau^{-2}\partial_{a}(h\rho^{ab}\pi_{b}^{w})  
\label{volvopitau}\\
& &
\ ( {\rm evolution \ equation \ of} \ \pi_{\tau})  \nonumber
\eea
where $H_{*}$ is defined by
\bea
& & \hspace{-0.6cm}
H_{*} = {1\over \tau} \rho_{a b}\rho_{c d}\pi^{a c}\pi^{b d}
+{1\over 4}\tau \rho^{a b} \rho^{c d}
(\partial_{R}\rho_{a c}) (\partial_{R}\rho_{b d})  \nonumber\\
& &  \hspace{0.2cm}
+\pi^{a c}\partial_{R}\rho_{a c}       
+  {1\over 2\tau}  \geq  {1\over 2 \tau} .   \label{trueham}     
\eea
{\it Einstein's constraint equations}
\bea
& &  \hspace{-0.4cm}
7.\ C=0  \Rightarrow  \pi_{\tau} = -H_{*} -2\pi_{R} \ ({\rm def.\ of} \ \pi_{\tau})                          
                             \label{pitau}\\
& & \hspace{-0.4cm}
8.\ C_{+}=0 \Rightarrow \pi_{R}=-\pi^{ab}\partial_{R}\rho_{ab}  \label{radi} \\
& & \ ({\rm def. \ of \ radial\ momentum})    \nonumber\\
& & \hspace{-0.4cm}
9.\ C_{a}=0  \Rightarrow \tau^{-1}{\pi}_{a}^{w}
= - \pi^{bc}\partial_{a}\rho_{bc}  +2\partial_{b}(\pi^{bc}\rho_{ac})   \nonumber\\
& & \hspace{-0.4cm}
-\tau\partial_{a}( H_{*}+ \pi_{R})    \ 
({\rm def. \ of \ angular \ momentum})   \label{angko}
\eea
{\it Superpotential} \  ${\ln}(-h)$
\bea
& &  
10.\ \partial_{\tau}  {\ln}(-h) =H_{*} - \tau^{-1}    \label{hstar}\\
& &  
11.\  -\partial_{R}  {\ln}(-h) =\pi_{R}    \label{pira}\\
& & 
12.\  -\partial_{a}  {\ln}(-h) = \tau^{-1}\pi_{a}^{w}   \label{piwa}
\eea
{\it Integrability conditions}
\bea
& & 
13.\ \partial_{R}(\tau^{-1}\pi_{a}^{w})=\partial_{a}\pi_{R}    \label{drtau}\\
& & 
14.\ \partial_{\tau}\pi_{R} = -\partial_{R}H_{*}  \label{partaur}\\
& & 
15.\  \partial_{\tau} (\tau^{-1}\pi_{a}^{w})=-\partial_{a}H_{*} \label{partautau}
\eea
The evolution equations of $\rho_{ab}$ and $\pi^{ab}$ can be found from the
reduced action  principle, 
\be
S_{*}=\int_{\Sigma_{3}}\!\!\!\! dRd^{2}Y
\{\pi^{ab}\partial_{\tau}\rho_{ab} -C_{*}^{(1)} \},
\ee
where $C_{*}^{(1)}$ is the restriction of  $C^{(1)}$ by  the coordinate 
conditions  $u=R$ and $y^{a}=Y^{a}$, 
\bea
& &
C^{(1)}_{*}:=C^{(1)}|_{u=R, y^{a}=Y^{a}}           \nonumber\\
& & 
=H_{*} +2\pi_{R} 
+2h\{ \tau R_{(2)}-{1\over 2}\tau^{-2}\rho^{ab}\pi_{a}^{w}
\pi_{b}^{w}              \nonumber\\
& & \hspace{0.2cm}
+\partial_{a}(\tau^{-1}\rho^{ab}\pi_{b}^{w}) \}.      \label{redaction}
\eea
Here $\pi_{R}$ is a total derivative given by (\ref{pira}). 
Variation of the reduced action  $S_{*}$ with respect to $h$ reproduces 
{\it  topological censorship constraint} (\ref{topo}), so the superpotential $h$ 
in (\ref{redaction}) is in fact a Lagrange  multiplier enforcing the constraint  (\ref{topo}), 
but this constraint {\it does} contribute to the following equations of motion:\\  
{\it Evolution equations of $\rho_{ab}$ and $\pi^{ab}$}
\bea
& &  \hspace{-3cm} 
16.  {\partial \rho_{ab}\over \partial \tau}=\int_{\Sigma_{3}} \!\!\!\! dR d^{2}Y {\delta C^{(1)}_{*}\over \delta \pi^{ab}}          \label{volvor}\\
& & \hspace{-3cm}
17.  {\partial \pi^{ab}\over \partial \tau}=-\int_{\Sigma_{3}} \!\! \!\! dR d^{2}Y {\delta C^{(1)}_{*}\over \delta \rho_{ab}}.             \label{volvop}
\eea
The spacetime metric in these privileged coordinates becomes
\be
ds^2 = -4h dR d\tau - 2h dR^2 +\tau\rho_{ab} dY^{a}  dY^{b}.
              \label{yoonspacetime}
\ee
On $\tau= {\rm constant}$ spacelike hypersurface $\Sigma_{3}$, the volume measure of $\Sigma_{3}$ is given by
\be
\sqrt{g}=\sqrt{-2h} \ \tau,
\ee
which increases monotonically in $\tau$,  as one finds that
\be
\partial_{\tau} {\ln} \sqrt{g}={1\over 2} H_{*}+ {1\over 2\tau} \geq 0,
\ee
where we used the equation (\ref{hstar}). 
Let us discuss some key features of above equations.\\
(i) First of all, let us mention that, without derivation, 
the whole set of the above equations are 
{\it identical to vacuum Einstein's equations}
$R_{AB}=0$ for spacetimes whose metric is given by (\ref{yoonspacetime}).
Thus, the whole procedure of the Hamiltonian reduction proposed in this Letter
respects the  general covariance, as it must, 
even though the final theory is written in the privileged coordinates. \\
(ii) The integral of (\ref{topo}) over a closed two-surface $N_{2}$
becomes
\be
\int_{N_{2}}\!\!\!\! d^{2}Y \tau^{-2}\rho^{ab}\pi_{a}^{w}\pi_{b}^{w}
= 16\pi (1-g) \geq 0,               \label{genus}
\ee
where $g$ is the genus of $N_{2}$. This identity states that,
as long as the out-going null hypersurface forms a congruence of
null geodesics which admits a cross section, the spatial topology of
that null hypersurface is either a two-sphere or a torus. This is an astonishingly
simple proof of {\it topological censorship}, as it
does not rely on assumptions such as global hyperbolicity,
asymptotic conditions, energy conditions, and so on, which are normally assumed
in the literature\cite{hawking72,fsw93,chru94,jacobson95,galloway96,gsww99}. \\
(iii) The spatial integral of $C^{(1)}_{*}$ defined in (\ref{redaction}) is the sought-for {\it physical} Hamiltonian of vacuum spacetimes. If we impose the topological censorship
constraint (\ref{topo}), the true Hamiltonian becomes
\be
\tilde{H}_{*}:=\int_{\Sigma_{3}} \!\! \! \! dR d^{2}Y H_{*} \geq 0,
\ee
which is {\it positive-definite} in the privileged coordinate.\\ 
(iv) The logarithm of the conformal factor in the $(\tau, R)$ subspace is the
{\it superpotential}  ${\ln} (-h)$, whose
gradients yield $(H_{*}, \pi_{R}, \tau^{-1}\pi_{a}^{w})$ through
(\ref{hstar}), (\ref{pira}), and (\ref{piwa}), respectively.
The superpotential\cite{kuchar71}  is a  local function of $x$, determined
by the line integral
\be
{\ln} {h(x) \over h(x_{0})}=  \int_{x_{0}, C}^{x} \!\!\!\!\!\!\!
\{ (H_{*}- \tau^{-1}) d\tau 
  -\pi_{R} dR 
  -\tau^{-1}\pi_{a}^{w}  dY^{a} \}            \label{potential}
\ee
along {\it any} contour $C$ from $x_{0}$ to $x$ in a given spacetime.\\
(v) The integrability conditions (\ref{drtau}), (\ref{partaur}), and (\ref{partautau})
are the consistency conditions,  which follow 
from the very definition of the superpotential ${\ln} (-h)$. \\
(vi) The momentum constraints are just  the {\it defining} equations of the radial  momentum density $\pi_{R}$ and angular momentum density 
$\tau^{-1}{\pi}_{a}^{w}$ in terms of true gravitational degrees of freedom
through (\ref{radi}) and (\ref{angko}). They do not restrict the theory in any way, 
so that the theory becomes {\it constraint-free}. The last two terms on the
right hand side of  (\ref{angko}), which are
total derivatives, represent coordinate effects as they do not contribute 
when integrated over a closed two-surface $N_{2}$.
If we define the total linear momentum $\tilde{\Pi}_{R}$ and total angular momentum
$\tilde{\Pi}_{a}$ as
\be
\tilde{\Pi}_{R}:=\int_{\Sigma_{3}} \!\! \!\! dR d^{2}Y  \pi_{R}, \hspace{0.2cm}
\tilde{\Pi}_{a}:=\int_{\Sigma_{3}} \!\! \!\! dR d^{2}Y \tau^{-1}{\pi}_{a}^{w},
\ee
then one finds that 
\bea
& & \tilde{\Pi}_{R}=-\int_{\Sigma_{3}} \!\! \!\! dR d^{2}Y 
\pi^{bc}\partial_{R}\rho_{bc},                     \label{totr}\\
& & \tilde{\Pi}_{a}
=-\int_{\Sigma_{3}} \!\! \!\! dR d^{2}Y
\pi^{bc}\partial_{a}  \rho_{bc}.     \label{tota}
\eea
These equations show that 
$\tilde{\Pi}_{R}$ and $\tilde{\Pi}_{a}$ are
the generating functions of translations of 
$\rho_{ab}$ and $\pi^{ab}$  along $\partial_{R}$ and
$\partial_{a}$. This justifies our interpretation of 
$\tilde{\Pi}_{R}$ and $\tilde{\Pi}_{a}$ 
as total linear and angular momentum
carried by physical degrees of freedom, respectively. 
Moreover, if suitable boundary conditions on $\rho_{ab}$ and $\pi^{ab}$ 
are assumed, then, 
by virtue of the integrability conditions (\ref{partaur}) and (\ref{partautau}),   
$\tilde{\Pi}_{R}$ and $\tilde{\Pi}_{a}$ are {\it conserved} in $\tau$ time!

\noindent
Details of this work and applications of this Hamiltonian formalism
to quantum theory  will be published in forthcoming papers. \\

{\bf ACKNOWLEDGEMENT}
This work is supported in part by Konkuk University (2011-A019-0035).

\end{document}